\documentclass{article}
\pdfoutput=1
\usepackage{spconf,amsmath,graphicx}
\usepackage{units}
\usepackage{balance}
\usepackage{units}
\usepackage{booktabs}
\usepackage{color}

\graphicspath{{./Plots/}}

\title{Enhancement of Coded Speech using a Mask-Based Post-filter }
\name{Srikanth Korse$^1$, Kishan Gupta$^{1,2}$ and Guillaume Fuchs$^{1,2}$}
\address{
  $^1$Fraunhofer IIS, Erlangen, Germany, srikanth.korse@iis.fraunhofer.de\\
  $^2$International Audio Laboratories, Friedrich-Alexander University (FAU), Erlangen, Germany 
  \sthanks{International Audio Laboratories is a joint institution between Fraunhofer IIS and Friedrich-Alexander University (FAU)}
  }
  
\begin{document}

\ninept
\maketitle

\begin{sloppy} 
\begin{abstract}
  The quality of speech codecs deteriorates at low bitrates due to high quantization noise. A post-filter is generally employed to enhance the quality of the coded speech. In this paper, a data-driven post-filter relying on masking in the time-frequency domain is proposed. A fully connected neural network (FCNN), a convolutional encoder-decoder (CED) network and a long short-term memory (LSTM) network are implemeted to estimate a real-valued mask per time-frequency bin. The proposed models were tested on the five lowest operating modes (6.65 kbps-15.85 kbps) of the Adaptive Multi-Rate Wideband codec (AMR-WB). Both objective and subjective evaluations confirm the enhancement of the coded speech and also show the superiority of the mask-based neural network system over a conventional heuristic post-filter used in the standard like ITU-T G.718. 
\end{abstract}


\section{Introduction} \label{sec:Introduction}
\paragraph*{}
State-of-the-art communication speech codecs such as 3GPP Adaptive Multi-Rate Wide-Band (AMR-WB)~\cite{AMRWB:2009alt} and 3GPP Enhanced Voice Services (EVS)~\cite{EVS:2014alt} use Code-Excited Linear Prediction (CELP) as a core coder for coding speech. CELP  consists of 3 essential parts: a short-term prediction using linear predictive coding (LPC), a long-term prediction (LTP) exploiting the fundamental frequency and the innovative codebook for modeling the residual of the predictions. At moderate to high bitrates, sufficient bits are assigned to the LPC coefficients, LTP parameters and the innovative codebook, thereby yielding sufficiently good to transparent quality. However, at low-bitrates, the majority of available bits are allocated to LPC and LTP parameters, and very few bits are left for the innovative codebook. This significantly affects the quality of coded speech at low bit rates.

To enhance the perceptual quality at low bitrates, modern speech codecs employ post-filters~\cite{chen1995adaptive,G718:2008,dietz2015overview,Tommy2015}. Post-processing the coded speech is conceptually similar to most speech enhancement techniques aiming to attenute or amplifiy frequency components with a bad or good signal-to-noise ratio, respectively. However, instead of modelling the noise or the speech, they rely mainly on prior knowledge and parameters of the coding scheme. Typical examples of post-processing are post-filters that emphasize the formants and the pitch structures of the coded speech by reusing information from LPC or LTP, respectively~\cite{chen1995adaptive,G718:2008}. It is also noteworthy that in~\cite{Das2018} a statistical model of the quantization noise of a very simplistic coding scheme is derived for designing an optimal post-filter in the log-magnitude domain.

In recent years, data-driven approaches for speech enhancement and dereverberation have been shown to outperform classical statistical signal processing approaches~\cite{Han2015,Zhao2016,Wang2014,Weninger2014}. They usually rely on data and make no assumptions about the signal or noise statistics. Among them, mask-based approaches~\cite{Wang2014, Weninger2014}, estimating either a real-valued ideal ratio mask (IRM), an ideal binary mask (IBM) or a complex-valued ideal ratio mask (cIRM) in the frequency domain, are especially efficient for denoising or inverse filtering tasks. These mask-based approaches have not been employed until now to enhance the quality of the coded speech and the authors hence investigate the benefit of using these approaches in the context of speech coding.   

Alternatively, the DNN can also learn the spectral mapping function between the spectral coefficients of noisy or reverberated speech and those of the clean speech~\cite{Han2015, Zhao2016}. This concept was recently adopted in~\cite{Zhao2018,Zhao2019} for enhancing  the coded speech in the cepstral domain. In order to reduce the complexity, the cepstrum is truncated and then fed as input to the neural network.

Recently, autoregressive models such as WaveNet~\cite{Oord2016} and LPCNet~\cite{Valin2019} have been employed to enhance the coded speech~\cite{Skoglund2019}. The high delay and complexity of these models prevent using them for real-time communication~\cite{G.114:2003}. Therefore, we do not consider these models for our evaluation. 

\subsection{Key Contribution of this Paper} \label{subsec:summary}
\begin{itemize}
\item The paper shows that a mask based post-filter in the spectral domain performs better than cepstral-domain post-filter (Cepstrum-CNN) as proposed in~\cite{Zhao2018,Zhao2019}.

\item FCNN, CED and LSTM structures were implemented to estimate a real-valued mask per time-frequency bin. This mask was then used to enhance the perceptual quality of the coded speech.

\item Among the three models proposed, CED performs better than FCNN and LSTM. Hence, CED was used for the final evaluation. 

\item The proposed model was trained on the coded speech obtained from the AMR-WB codec at bitrate \unit[6.65]{kbps}. The trained model was then tested on all bitrates ranging from \unit[6.65]{} to \unit[15.85]{kbps}.

\item Robustness of the trained model was validated by testing on a completely different database.

\item The proposed model is also compared with the heuristic post-filter adopted in G.718~\cite{G718:2008}.
\end{itemize}

\section{Problem Formulation}\label{sec:Prob_Form}
\subsection{Oracle Experiments} \label{subsec:Oracle_Exp}
From a simplistic mathematical point of view, one can describe the coded speech $\tilde{x}(n)$ as:
\begin{equation}\label{equ_quantNoise}
\tilde{x}(n) = x(n) + \delta(n)
\end{equation} 
where $x(n)$ is the input speech to the codec and $\delta(n)$ is the quantization noise. 
The quantization noise $\delta(n)$ is correlated with the input speech since CELP uses a perceptual model during the quantization process. The correlation of quantization noise with the input speech makes our post-filtering problem unique to speech enhancement problem which usually assumes the noise to be uncorrelated. 

In order to reduce the quantization noise,  we estimate a real-valued mask per time-frequency bin and multiply this mask with the corresponding spectral magnitude of the coded speech as shown:

\begin{equation}\label{equ_Mask}
|\hat{X}(k,n)| = M(k,n) \cdot |\tilde{X}(k,n)|,
\end{equation}
where $M(k,n)$ is the real-valued mask, $|\tilde{X}(k,n)|$ is the magnitude spectrum of the coded speech, $|\hat{X}(k,n)|$ is the magnitude spectrum of the enhanced speech, $k$ is the frequency index and $n$ is the time index. 

If the mask is ideal, we can get back the clean speech magnitude spectrum from the coded speech magnitude spectrum. The definition of the ideal ratio mask (IRM) is given by:   
  
\begin{equation}\label{equ_IRM}
\text{IRM}(k,n) = \frac{|X(k,n)|}{|\tilde{X}(k,n)|+\gamma},
\end{equation}
where $|X(k,n)|$ is the magnitude spectrum of the clean speech and $\gamma$ is a very small constant factor to prevent division by zero. Since the magnitude values lie in the range $0$ to $\infty$, the values of IRM is positive and unbounded.

Table~\ref{tab:maskValueDistr_AMR} shows the distribution of real-valued mask per time-frequency bin in different threshold regions at lowest three birates of AMR-WB. These mask values were computed using (\ref{equ_IRM}). From the Table~\ref{tab:maskValueDistr_AMR}, it can be concluded that, since most the mask values lie in the region between $0$ and $5$, we cannot afford to bound the mask value to 1 as usually done in the conventional speech enhancement techniques. 

\begin{table}[t]
  \centering
   \resizebox{0.4\textwidth}{!}
   {
  \begin{tabular}{ cccc  }
  \toprule
  \textbf{Mask Thresholds} & \textbf{6.65kbps} & \textbf{8.85kbps} & \textbf{12.65kbps} \\
  \midrule
  $[0,1]$ & 38.94\% & 41.00\% & 44.09\% \\
  $(1,2]$ &  31.19\% & 33.44\% & 36.20\%  \\
  $(2,5]$ &  21.40\% & 18.69\% & 14.66\% \\
  $(5,\infty]$ &  8.46\% & 6.87\% & 5.05\% \\
  \bottomrule
  \end{tabular}
  } 
  \caption{Percentage of real-valued mask in different threshold regions measured at lowest three birates of AMR-WB. }
  \label{tab:maskValueDistr_AMR} 
\end{table}

To find out the ideal bounding value and also the performance of the mask-based technique over cepstral methods, oracle experiments were performed at \unit[6.65]{kbps} and \unit[8.85]{kbps}. The oracle mask was computed using (\ref{equ_IRM}) and then was bounded to values 1, 2, 4 and 10. The oracle cepstrum was computed as follows: First, cepstrum of length 512 was obtained from both coded and clean speech. Second, the first 64 cepstral values obtained from the coded speech were replaced by the clean speech cepstral values.   
  
\begin{figure}[t]
  \centering
  \includegraphics[width=0.82\linewidth]{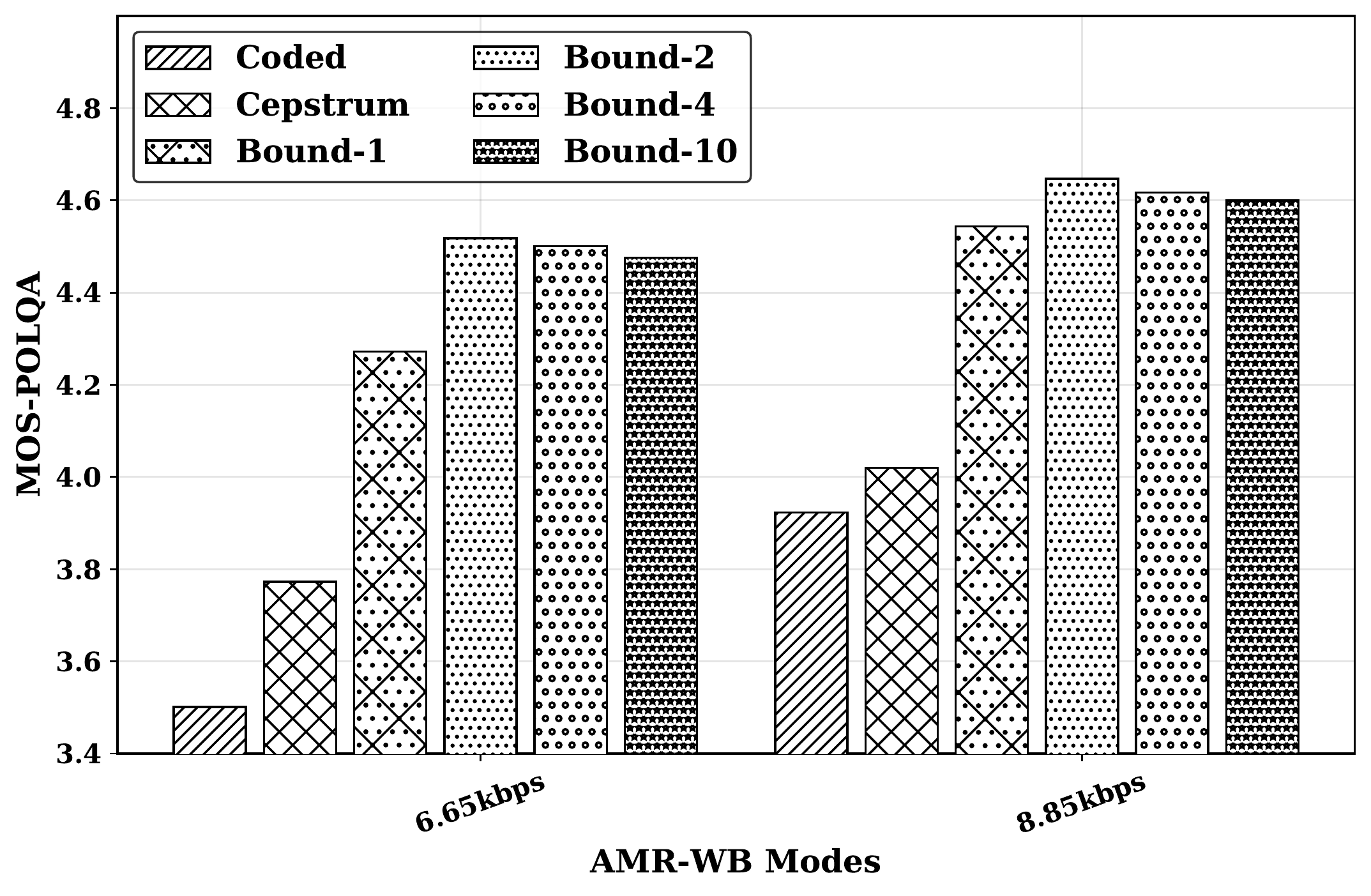}
  \caption{Average POLQA scores evaluating the oracle experiment}
  \label{fig:Oracle}
\end{figure}

Fig.~\ref{fig:Oracle} compares the average Perceptual Objective Listening Quality Assessment (POLQA)~\cite{POLQA} scores between coded, oracle cepstrum and oracle mask with different bounding values at bitrates \unit[6.65]{kbps} and \unit[8.85]{kbps}. It can be clearly observed that, in the oracle case, spectral mask-based methods outperform the cepstrum method. With regards to the upper bound, it can be observed that the bound values greater than 1 perform significantly better than bound 1. Among the bound values 2, 4 and 10, there is no siginificant difference. This motivated us to bound our mask to 2 in further experiments. 

The above observations in the oracle experiments motivate the adoption of mask-based approach in the spectral domain for enhancing the perceptual quality of the coded speech. 

\subsection{Modified Signal Approximation} \label{subsec:Mod_SA}
We train the neural network using modified signal approximation (mod-SA). The major differences between mod-SA and traditional signal approximation (SA) defined in~\cite{Weninger2014} are as follows:

\begin{itemize}
\item Instead of computing the IRM and bounding it to 2, the modified mask $\tilde{M}(k,n)$ is computed as:

\begin{equation}\label{modified_mask}
\tilde{M}(k,n) =
  \begin{cases}
    \text{IRM}(k,n)       & \quad \text{if} \quad \text{IRM}(k,n) \leq \alpha\\
    \rho  & \quad \text{if} \quad \text{IRM}(k,n) > \alpha
  \end{cases}
\end{equation}

\noindent where $0 \leq \rho \leq \alpha$ and $\text{IRM}(k,n)$ is computed using (\ref{equ_IRM}). For our experiments, we choose  $\alpha$ as 2 and $\rho$ as 1. In other words, when the IRM is greater than bound 2, the coded spectral magnitude is kept unchanged. 

\item Since the mask was modified, the target is also modified and the modified target $|\bar{X}(k,n)|$ is given by: 
\begin{equation}\label{equ_ModTarget}
|\bar{X}(k,n)| = \tilde{M}(k,n) \cdot |\tilde{X}(k,n)|,
\end{equation}

\item The mean square error (MSE) is computed between the modified target and enhanced speech in the log-magnitude domain instead of the magnitude domain.
\end{itemize}   

The above modifications are essential to get a generalized model that works on the higher bitrates despite being trained only on the lowest one. If the objective is to have different models for different bitrates, the authors would advise to use then the unmodified target $|X(k,n)|$ for training.  

\section{Experimental Setup}\label{sec:Expt_Setup}
Our proposed post-filter computes the short-time Fourier transform (STFT) of frames of 32 ms with 50\% overlap (16 ms) at 16 kHz sampling rate resulting in 257 frequency bins. The square root of the Hann window is used as analysis and synthesis window. Only bandwidth up to 6.4 kHz (205 frequency bins) was processed with DNN and the frequency region  between 6.4 to 7 kHz was left unprocessed. Since speech has temporal dependency, past frames were used as context frames i.e. input to the DNN were past frames plus the current frame. The input to the DNN was normalized log-magnitude since magnitude values have a higher dynamic range compared to log-magnitude. The output of the DNN is a real-valued mask that lies in the range $0$ to $2$. The real-valued mask is then multiplied with the coded magnitude to obtain the enhanced magnitude. 

The three proposed networks are explained below in detail:

\begin{itemize}
\item \textbf{FCNN}: Three past frames and the current frame are concatenated and provided as input to the FCNN. The size of the input was 820. It has two hidden layers with 1024 units. Each hidden layer consists of Rectified Linear Units (ReLU) as activation functions along with batch normalization and a dropout of 0.2. The output layer consists of 205 units. 

\item \textbf{LSTM}: The LSTM network consists of two LSTM layers with 400 and 205 units, respectively, with 10 time steps (9 past frames and current frame). A Dropout of 0.1 and recurrent dropout of 0.2 was used. Only the last time step of the second LSTM unit was given as input to the output layer. 

\item \textbf{CED}: An encoder-decoder architecture-based CNN is implemented as shown in Table \ref{tab:CNN_architecture}. The input to the CED is 6 time steps (5 past frames and current frame). Each layer of CNN uses batch normalization and ELU (Exponential Linear Unit) activation function. Skip connections are used between encoder and decoder. Required zero-padding is done in the time frame to match the \texttt{frequencyBins} dimensions for skip connections.

\begin{table}[t]
    \setlength{\arrayrulewidth}{0.2mm}
    \centering
    \resizebox{0.40\textwidth}{!}{
          \begin{tabular}{ |c|c|c|c|  }
          \hline
          \textbf{Layer name} & \textbf{Input} & \textbf{Hyperparameter} & \textbf{Output} \\
          \hline
          Reshape & $6 \times 205$ & - & $1 \times 6 \times 205$ \\
          \hline
          Conv2d\textunderscore1 & 
          $1 \times 6 \times 205$ & $2 \times 3$, (1,2), 16  & $16 \times 5 \times 102$ \\
          \hline
          Conv2d\textunderscore2 & $16 \times 5 \times 102$ & $2 \times 3$, (1,2), 32  & $32 \times 4 \times 50 $ \\
          \hline
          Conv2d\textunderscore3 & $32 \times 4 \times 50 $ & $2 \times 3$, (1,2), 64  & $64 \times 3 \times 24 $ \\
          \hline
          Conv2d\textunderscore4 &  $64 \times 3 \times 24 $ & $2 \times 3$, (1,2), 128  & $128 \times 2 \times 11 $ \\
          \hline
          Deconv2d\textunderscore1 & $128 \times 2 \times 11 $ & $2 \times 3$, (1,2), 64  & $64 \times 3 \times 23 $ \\
          \hline
          Deconv2d\textunderscore2 & $128 \times 3 \times 24 $ & $2 \times 3$, (1,2), 32  & $32 \times 4 \times 49 $ \\
          \hline
          Deconv2d\textunderscore3 & $64 \times 3 \times 50 $ & $2 \times 3$, (1,2), 16  & $16 \times 5 \times 101 $ \\
          \hline
          Deconv2d\textunderscore4 & $32 \times 5 \times 102 $ & $2 \times 3$, (1,2), 1  & $1 \times 6 \times 205 $ \\
          \hline
          Conv2d\textunderscore5 & $1 \times 6 \times 205 $ & $6 \times 1$, (1,1), 1  & $1 \times 1 \times 205 $ \\
          \hline
          Flatten & $1 \times 6 \times 205 $ & - & $1 \times 205 $ \\
          \hline
          \end{tabular} }
    \caption{Architecture of our proposed CED. The input and output size is given as \texttt{featureMaps} $\times$ \texttt{timesteps} $\times$ \texttt{frequencyBins}. The hyperparameter is indicated as \texttt{kernelsize, strides, outchannels}. }
    \label{tab:CNN_architecture} 
\end{table}

\end{itemize}

For all the models, the output layer consists of sigmoid units scaled with factor 2. All the models were trained with the ADAM optimizer~\cite{Kingma2014} with a learning rate of 0.001 and a batch size of 32. Instead of using a fixed number of epochs, training was done till convergence using early stopping.  The implementation of the reference Cepstrum-CNN is the same as proposed in~\cite{Zhao2018}. It is to be noted that the Cepstrum-CNN do not use any context frames. 

\begin{table}[t]
    \setlength{\arrayrulewidth}{0.2mm}
    \centering
    \resizebox{0.40\textwidth}{!}{
          \begin{tabular}{ |c|c|c|  }
          \hline
          \textbf{Network Architecture} & \textbf{Number of Parameters} & \textbf{Frame Size}\\
          \hline
          FCNN & $2108621$ & $32$ms\\
          LSTM & $1468120$ & $32$ms\\
          CED & $\textbf{147292}$ & $32$ms\\
          Cepstrum-CNN & $419805$ & $\textbf{20ms}$\\
          \hline
          \end{tabular} }
    \caption{Comparison of the number of parameters in different network architectures. }
    \label{tab:architecture_params} 
\end{table}
Table~\ref{tab:architecture_params} shows the number of parameters and frame sizes for each of the compared networks. Our proposed CED model has the smallest number of parameters in comparison to all other models. On other hand, our model operates on \unit[32]{ms} frames compared to Cepstrum-CNN which operates on \unit[20]{ms}. This results in \unit[16]{ms} delay for our model instead of \unit[10]{ms} delay for Cepstrum-CNN, which is still acceptable for real time communication. 

The input speech signals were first filtered using the P.341 filter (cutoff frequency of 7kHz) ~\cite{G191:2005} and then the active speech level was adjusted to -26 dBov~\cite{P56:2011} before coding with AMR-WB. The coded speech signals were also filtered using P.341 filter with same 7 kHz cutoff frequency before testing. For objective assessment, we used POLQA, while for subjective assessment, we followed the methdology MUltiple Stimuli with Hidden Reference and Anchor (MUSHRA)~\cite{MUSHRA}. 

The time-domain enhanced speech was obtained using the inverse short-time Fourier transform (iSTFT) and a synthesis window before the ovelap and add step. iSTFT made use of the phase of the coded speech without any processing.  

\section{Experiments and Results}\label{sec:Expt_Results}
\begin{figure}[t]
  \centering
  \includegraphics[width=0.80\linewidth]{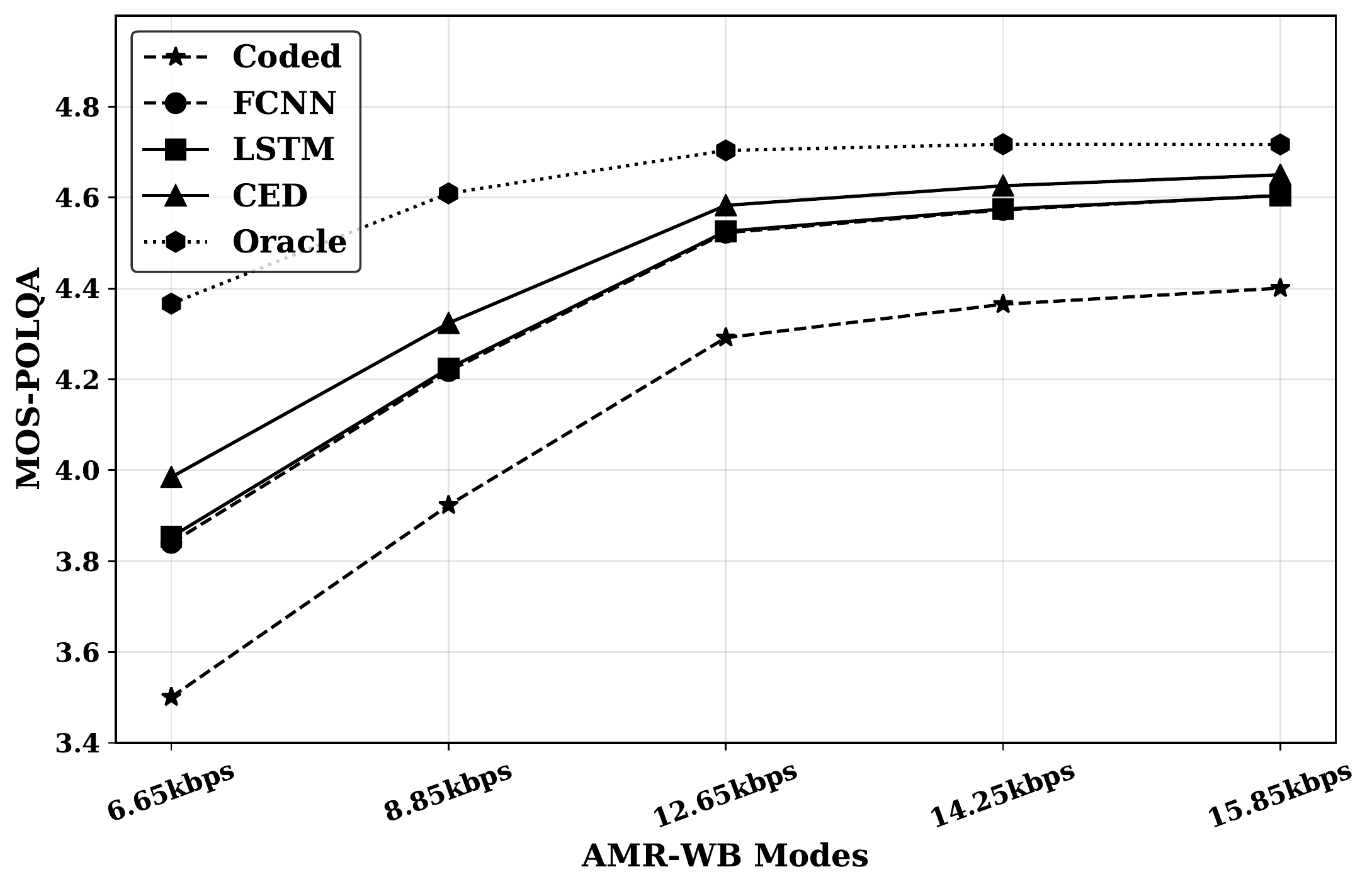}
  \caption{POLQA scores evaluating the performance of the FCNN, LSTM and CED architectures using the NTT test set.}
  \vspace{-0.25cm}
  \label{fig:ConextFrames}
\end{figure}

\begin{figure}[t]
  \centering
  \includegraphics[width=0.80\linewidth]{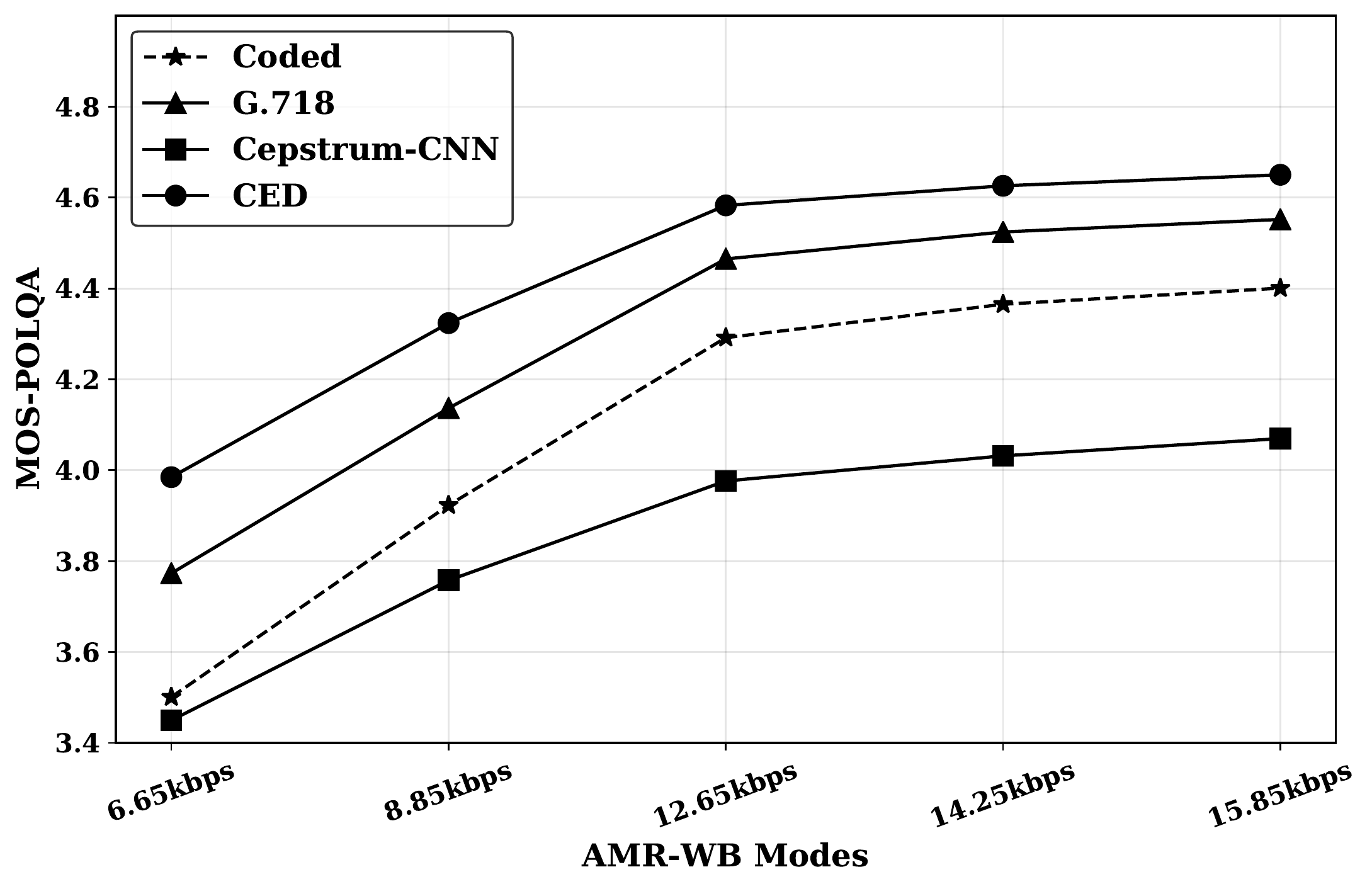}
  \caption{POLQA scores evaluating the performance of the Cepstrum-CNN, CED and G.718 using the NTT test set.}
  \vspace{-0.25cm}
  \label{fig:Evaluation}
\end{figure} 

For training, we used the NTT-AT~\cite{nttdb:2012} database. The files were downsampled to 16 kHz and a passive mono downmix was obtained from the stereo files. Out of 3690 files, 3612 files were used for training, 198 files were used for validation and 150 files were used for testing. The database was split in such a way that the distribution in terms of male and female speakers and languages was balanced in the training, validation and test set. All the files were preprocessed as explained in Section \ref{sec:Expt_Setup}.  

Fig. \ref{fig:ConextFrames} compares the POLQA scores of the three proposed architectures (FCNN, CED and LSTM). Among the proposed architectures, LSTM and FCNN have the same performance while CED consistently performs better than the two others across all bitrates. Hence, for further evaluation, we only consider the CED architecture. The oracle mask used for comparison was obtained as explained in Section \ref{subsec:Mod_SA}. 
 
Fig. \ref{fig:Evaluation} compares the POLQA scores of the proposed CED model with the post-filter used in G.718 and the Cepstrum-CNN model. Our proposed CED model improves the perceptual quality of the coded speech and is consistently better than G.718 at all bitrates. The Cepstrum-CNN fails to improve the perceptual quality of the coded speech at all bitrates.

\begin{figure}[t]
  \centering
  \includegraphics[width=0.80\linewidth]{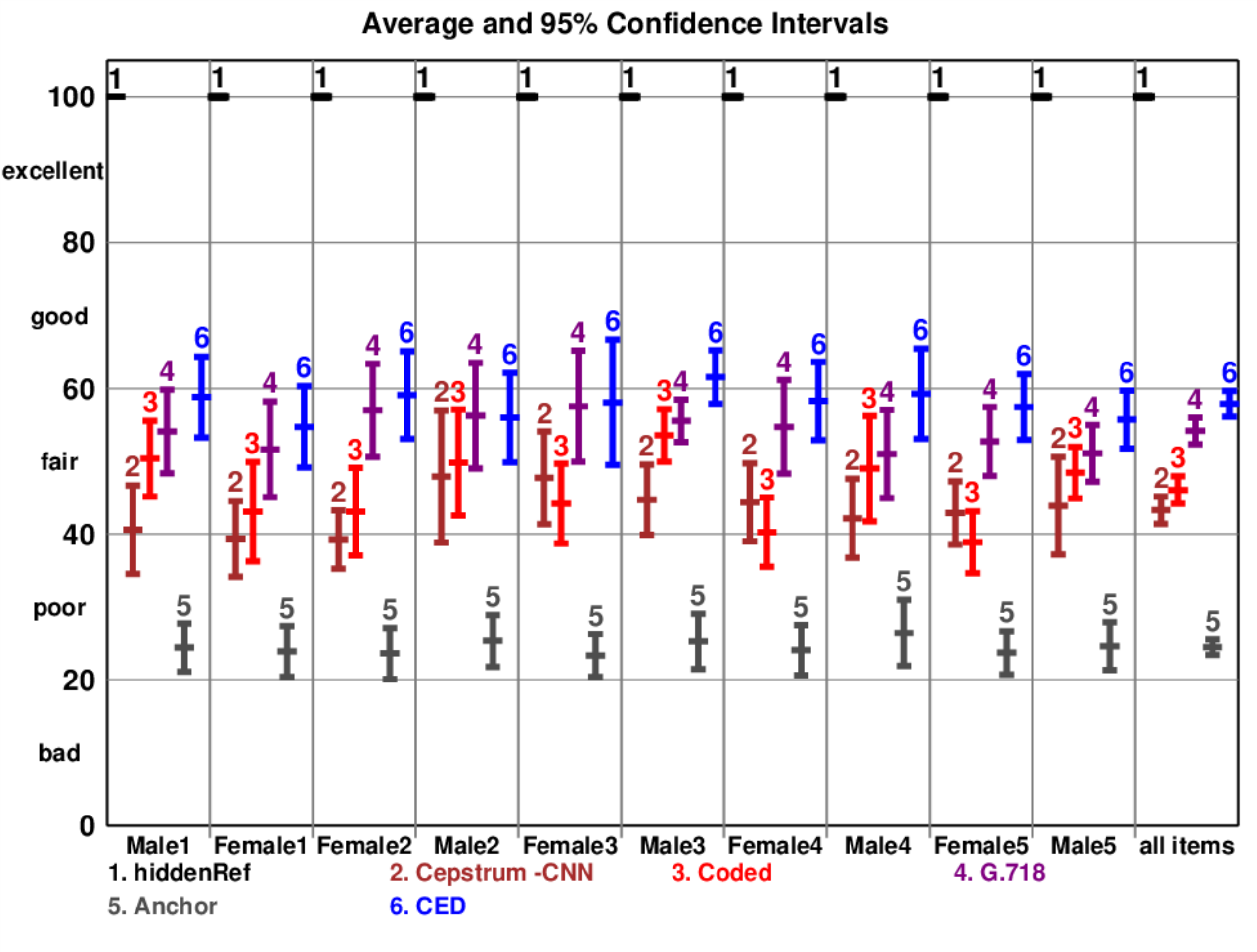}
  \caption{Average MUSHRA scores of 11 listeners at \unit[6.65]{kbps}.}
  \vspace{-0.25cm}
  \label{fig:LT_Mode0}
\end{figure}

The POLQA scores are consistent with the MUSHRA scores as shown in Fig. \ref{fig:LT_Mode0} and Fig. \ref{fig:LT_Mode2} which compare the proposed CED model with G.718 and the Cepstrum-CNN at bitrates 6.65 and 12.65 kbps, respectively. Both listening tests involve 11 expert listeners. The listening tests were conducted in the listening test rooms which were isolated from the outside noise. STAX headphones were used for the listening tests. 

\begin{figure}[t]
  \centering
  \includegraphics[width=0.80\linewidth]{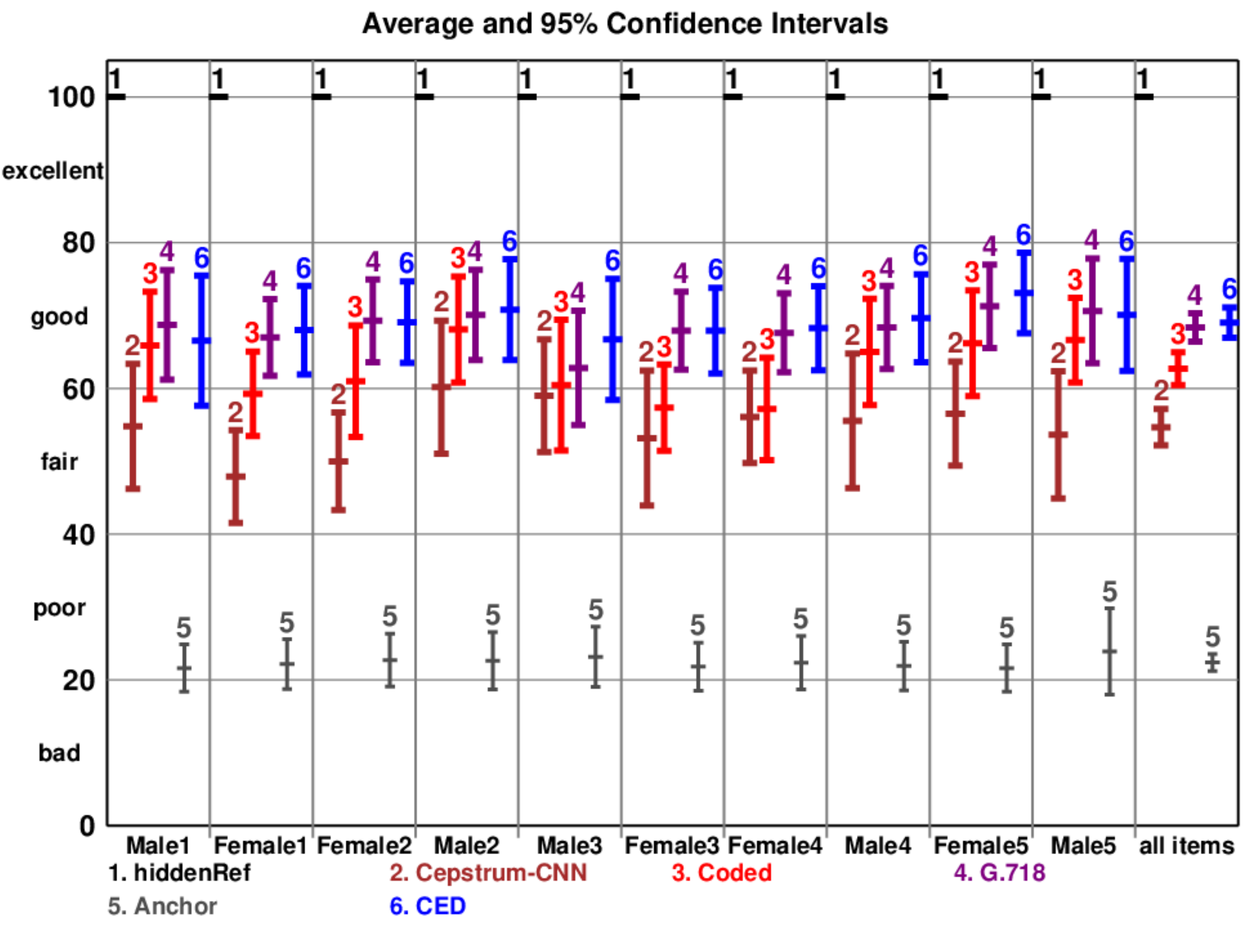}
  \caption{Average MUSHRA scores of 11 listeners at \unit[12.65]{kbps}. }
  \vspace{-0.2cm}
  \label{fig:LT_Mode2}
\end{figure}

At \unit[6.65]{kbps}, our proposed CED model sucessfully enhances the coded speech and is also better than the G.718. Compared to the coded speech, the CED model gains around 12 MUSHRA points and 0.5 POLQA mean opinion score (MOS). At \unit[12.65]{kbps}, the benefit of our proposed CED model is less and is close to G.718 post-filter. This is primarily because the quality of the coded speech lies already in the good range. 

The Cepstrum-CNN model fails to enhance the quality of coded speech at both \unit[6.65]{kbps} and \unit[12.65]{kbps}. This is due to the following reasons:

\begin{itemize}
\item Since only the first cepstral coefficients are enhanced, it affects only the spectral envelope and not the spectral fine structures. 
\item The Cepstrum-CNN fails to supress the noise between harmonics compared to our proposed CED model or the G.718 post-filter.  
\item At high frequencies, the Cepstrum-CNN is closer to the target signal in terms of energy, but at the price of amplifying the coded artefacts. 
\item In addition, at high frequencies, our proposed CED model has better capability of restoring the harmonic structure lost due to high quantization noise compared to the Cepstrum-CNN. 
\end{itemize}

In order to test the model on completely unseen data, we performed a cross-database validation test. Fig  \ref{fig:CrossDatabaseTesting} shows the performance of all the models using the test set of TIMIT database~\cite{garofolo1993timit}. The behavior is similar to the test set of the NTT database, thus validating that the proposed model works well on completely unknown data.

\begin{figure}[t]
  \centering
  \includegraphics[width=0.80\linewidth]{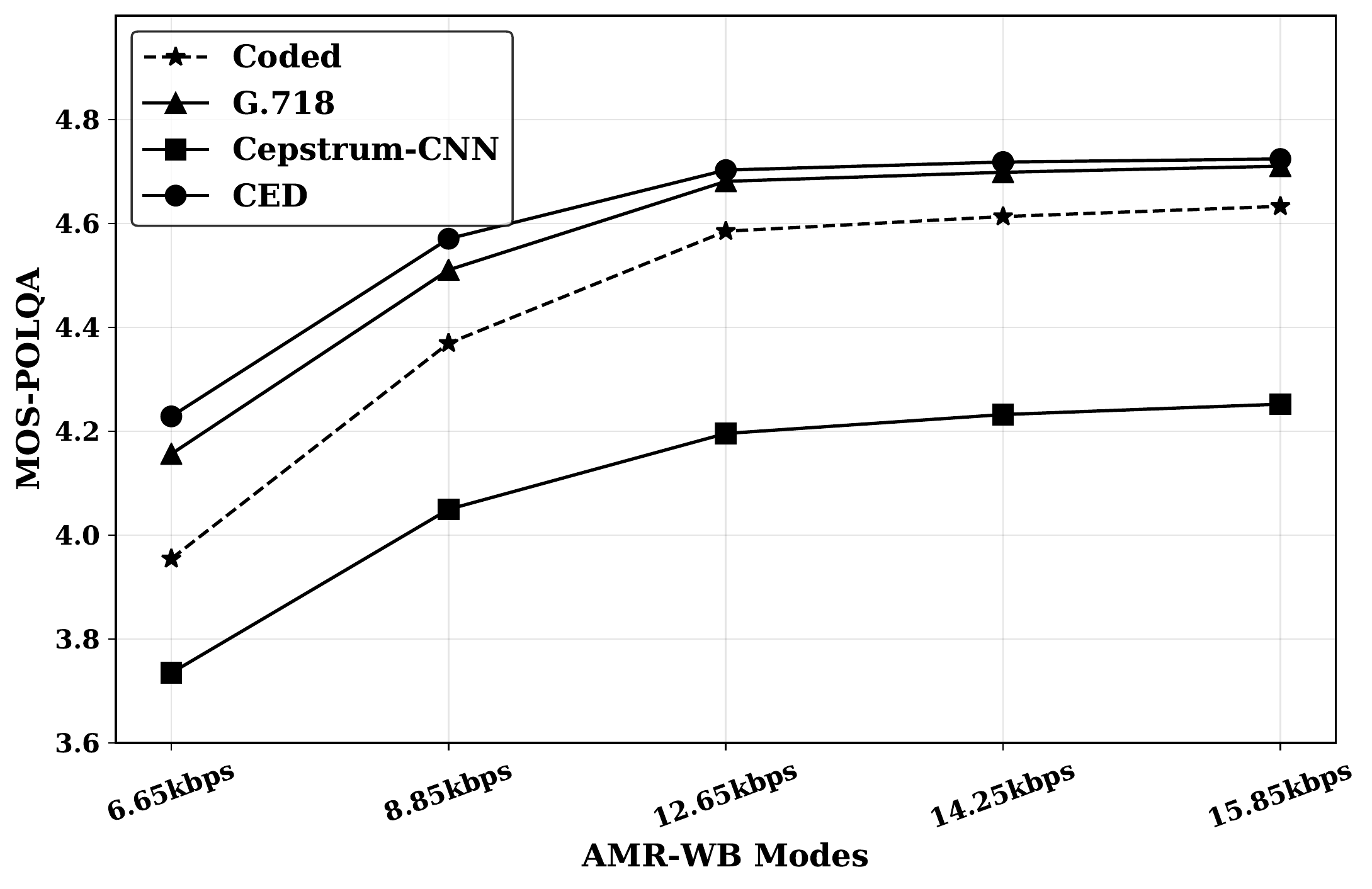}
  \caption{POLQA scores evaluating the performance of Cepstrum-CNN, CED and G.718 using the TIMIT test set }
  \label{fig:CrossDatabaseTesting}
\end{figure}

\section{Conclusion}\label{sec:Discussions_Conclusions}
A convolutional encoder-decoder (CED) based post-filter that estimates a real-valued mask per time frequency bin is proposed to enhance the quality of the coded speech. It is shown that modified-signal approximation is necessary to train a generalized model that works well at higher bitrates despite being trained on the lowest one. Based on POLQA and MUSHRA scores, it was confirmed that real-valued mask based post-filter based on data driven approach that makes no assumption about signal or noise characteristics can successfully enhance the quality of the coded speech. The benefits are higher at low bitrates and as the bitrate increases, the benefits observed are smaller. Cross-database testing also confirmed the robustness of our proposed CED model.

\bibliographystyle{IEEEtran}
\bibliography{refs19}

\end{sloppy}
\end{document}